\documentclass[preprint,preprintnumbers,nofootinbib,aps,10pt]{revtex4-1}
 \usepackage{amsmath,amssymb,bm,epsfig}
 \usepackage{color}
 \usepackage{natbib}
 \usepackage{hyperref}
 \usepackage{ulem}
 \usepackage{graphicx}
 \usepackage{xcolor,colortbl}
 \usepackage{pifont}
\def \beq{\begin{equation}}
\def \eeq{\end{equation}}
\def \beqa{\begin{eqnarray}}
\def \eeqa{\end{eqnarray}}
\def \l{\left(}
\def \r{\right)}

\newcommand{\nn}{\nonumber}

\usepackage{xspace}

\newcommand{\sNN}{\sqrt{s_{\rm NN}}}
\begin{document}

\title{Production of Light Nuclei in Heavy Ion Collisions Within Multiple 
Freezeout Scenario}

 \author{Sandeep Chatterjee}
 \email{chatterjee.sandeep@niser.ac.in}
 \affiliation{School of Physical Sciences, 
  National Institute of Science Education and Research, Bhubaneshwar, 
  751005, India}
 \author{Bedangadas Mohanty}
 \email{bedanga@niser.ac.in}
 \affiliation{School of Physical Sciences, 
  National Institute of Science Education and Research, Bhubaneshwar, 
  751005, India}


\begin{abstract}

We discuss the production of light nuclei in heavy ion collisions within 
a multiple freezeout scenario. Thermal parameters extracted from the 
fits to the observed hadron yields are used to predict the multiplicities 
of light nuclei. Ratios of strange to non strange nuclei are found to be 
most sensitive to the details of the chemical freezeout. The well known 
disagreement between data of $^3_\Lambda\text{H/}^3\text{He}$ and 
$\overline{^3_\Lambda\text{H/}^3\text{He}}$ at $\sqrt{s_{NN}}=200$ GeV and 
models based on thermal as well as simple coalescence using a single 
chemical freezeout surface goes away when we let the strange and non strange 
hadrons freezeout at separate surfaces. At the LHC energy of $\sNN=2700$ GeV, 
multiple freezeout scenario within a thermal model provides a consistent 
framework to describe the yields of all measured hadrons and nuclei.\\
PACS numbers:
\end{abstract}
\maketitle

Hadron resonance gas models have been traditionally employed to understand 
the production of hadrons in heavy ion collisions across beam energies 
varying by several orders of magnitudes. This is done with a few thermal 
parameters like volume $V$, temperature $T$ and chemical potentials $\mu_B$, 
$\mu_Q$ and $\mu_S$ corresponding to the conserved charges (in QCD) baryon 
number $B$, electric charge $Q$ and strangeness $S$ respectively that describe 
the thermal state of the fireball at the time of chemical freezeout (CFO). 
By comparing the hadron yields to the thermal model predictions, the 
thermodynamic state of the fireball can be deduced at the time of CFO. The 
standard practice has been to assume a single freezeout surface for all hadrons 
which we call here 1CFO~\cite{BraunMunzinger:1995bp,Yen:1998pa,Becattini:2000jw}. 
The recent LHC data on hadron yields at $\sNN=2700$ GeV, particularly ratios 
of strange to non strange baryons like $\Lambda/$p, $\Xi/$p and $\Omega/$p could 
not be explained by thermal models with 1CFO~\cite{Stachel:2013zma}. This has 
initiated new efforts in understanding the hadrochemistry at the time of chemical 
freezeout. In Refs.~\cite{Steinheimer:2012rd,Becattini:2012xb,Becattini:2012sq}, 
the effects of late stage hadronic inelastic scattering was computed that mainly 
led to proton-antiproton annihilation resulting in better agreement with data. 
There has been suggestion to introduce non-equilibrium phase space factors for 
light and strange quarks~\cite{Petran:2013lja} which also lead to agreement with 
data. In Refs.~\cite{Chatterjee:2013yga,Bugaev:2013sfa} it was argued on the basis 
of hadrochemistry that non strange and strange hadrons are expected to freezeout 
at different times (2CFO). It was demonstrated that thermal model fits to hadron 
yields based on 2CFO improve considerably specially at the LHC 
energy~\cite{Chatterjee:2013yga}. From the microscopic point of view, because of 
the varying hadronic cross sections among the various hadrons in the medium, a 
sequential freezeout is expected to happen. In this paper we study the production 
of light nuclei in heavy ion collisions within 2CFO.

Recently STAR has reported the first observation of antihypernuclei at $\sNN=200$ 
GeV~\cite{Abelev:2010rv}. While hypernuclei has been observered earlier, 
antihypernuclei have been elsusive for a long time. Previous studies have found 
that at $\sNN=200$ GeV, thermal and simple coalescence models with 1CFO describe 
antinuclei to nuclei ratios as well as ratios of non strange or strange nuclei like 
$\overline{^3\text{He}}/^3\text{He}$, $\overline{^3_\Lambda\text{H}}/^3_\Lambda\text{H}$, 
${}^3\text{He}/$H, d/p etc. On the other hand in the case of mixed ratios 
(strange to non strange nuclei ratios) like $^3_\Lambda\text{H/}^3\text{He}$ and 
$\overline{^3_\Lambda\text{H}}/\overline{^3\text{He}}$, 1CFO is found to 
underpredict at the top STAR energy of $\sNN=200$ GeV~\cite{Andronic:2010qu,
Cleymans:2011pe}. In this paper we investigate the production of nuclei yields within 
2CFO and show that in thermal as well as simple coalescence models, strange to non 
strange nuclei ratios demonstrate the necessity of 2CFO model over 1CFO.

The production of light nuclei in heavy ion collisions has been successfully studied 
within two phenomenological models with very different mechanism. Thermal models that 
successfully describe the hadron yields are even found to explain nuclei yields based 
on early stage production of the nuclei at the CFO surface along with the other hadrons. 
It was first pointed out in Refs.~\cite{Mekjian:1978zz,Siemens:1979dz} that light nuclei 
could also equilibrate chemically along with other hadrons in the expanding fireball 
produced in heavy ion collisions. Thus, there has been analysis of light nuclei yields 
based on thermal models with 1CFO which has been quite successful in describing their 
multiplicity~\cite{BraunMunzinger:1994iq,BraunMunzinger:2001mh, Andronic:2010qu,
Cleymans:2011pe}. However, a detailed study of the effect of the small binding energy 
(compared to the fireball temperature) and reaction kinetics can not be investigated 
in such models. Secondly, coalescence models where there is late stage production by 
recombination of the constituent hadrons near the kinetic freezeout (KFO) surface can 
also describe the production of light nuclei. These require proper ways to incorporate 
correlations between the constituent hadrons in their phase spaces so that they could 
coalesce to form nuclei~\cite{Butler:1963pp,Schwarzschild:1963zz,Gutbrod:1988gt,
Sato:1981ez,Scheibl:1998tk,Nagle:1996vp,Llope:1995zz,Danielewicz:1991dh,Dover:1991zn,
Csernai:1986qf,Gyulassy:1982pe,Remler:1981du,Zhang:2009ba,Steinheimer:2012tb}. In both 
the scenarios, the CFO surface plays a crucial role. While in thermal models, the details 
of the CFO enter  as the thermal parameters of the light nuclei themselves, in the case 
of the colaesence viewpoint they enter through the thermal parameters of the constituents. 
The aim of this work is to demonstrate the role played by the CFO scheme that one employs 
in determining the light nuclei yields both in thermal as well as coalescence models. For
our purpose it is sufficient to restrict to the simplest version of the coalescence model 
where effects due to non trivial correlations in the phase space of the hadrons that 
coalesce to form nuclei are ignored. For the same reason, in the simple coalescence model 
we will only deal with ratios of nuclei while in the thermal model we will compute the 
yields in addition to the ratios of nuclei.

We compute the thermal as well as coalescence model predictions for different ratios of 
nuclei at different $\sNN$ using the thermal parameters extracted at the corresponding 
energies from fits to hadron yields within 2CFO~\cite{Chatterjee:2013yga}. Particularly 
at $\sNN=200$ GeV, we study the sensitivity of the 2CFO parameters on light nuclei yields 
by including the available data on light nuclei in our fit. We also find a consistent 
description of all the measured hadron and nuclei yields at $\sNN=2700$ GeV. The paper is 
organised as follows: In Section \ref{sec.Model} we briefly discuss details of the 2CFO 
scheme as implemented in the thermal and coalescence models. In Section \ref{sec.results} 
we will present the model prediction for ratios of nuclei as obtained in the 2CFO scheme 
and compare them with that of 1CFO. We point out that amongst different nuclei ratios, 
mixed ratios (strange to non strange nuclei) are particularly suitable to probe and 
discriminate between different CFO schemes. At $\sNN=200$ GeV, very good agreement between 
model prediction and data is found for the mixed ratios $^3_\Lambda\text{H/}^3\text{He}$ 
and $\overline{^3_\Lambda\text{H/}^3\text{He}}$ within the 2CFO scheme. We find that 
thermal model with 2CFO provides a consistent picture that describes the yield of all 
measured hadron and nuclei yields at the LHC Pb-Pb collision at $\sNN=2.76$ TeV. Finally 
in Section \ref{sec.conc} we summarise and conclude.

\section{2CFO SCHEME}
\label{sec.Model}
\subsection{THERMAL MODEL}
\label{subsec.2CFOThermal}
The ideal hadron resonance gas (HRG) partition function $Z$ in the grand canonical
ensemble at the time of CFO at a particular beam energy $\sNN$  is given 
as 
\beq
 \log\left[Z\l\sNN\r\right]=\sum_i\log\left[Z_i\l T_i\l\sNN\r,
 \mu_i\l\sNN\r, V_i\l\sNN\r \r\right]
 \label{eq.Z}
\eeq
where $Z_i$ is the partition function of the $i$th hadron species and $T_i$, $\mu_i$ and 
$V_i$ are its relevant thermal parameters at the time of CFO. Thus the 
primordial yield $N_i^{\text{p}}$ of the $i$th hadron is given by
\beqa
 N_i^{\text{p}}&=&\frac{\partial}{\partial \l\frac{\mu_i}{T_i}\r}\log\left[Z\right]\nn\\
 &=&\frac{V_iT_i}{\pi^2}g_im_i^2\sum_{l=1}^{\infty}\l-a\r^{l+1}l^{-1}
  K_2\l lm_i/T_i\r\times\exp\l l\l B_i{\mu_B}_i+Q_i{\mu_Q}_i+S_i{\mu_S}_i\r/T_i\r
 \label{eq.primyield}
\eeqa
where  $a=-1$ for bosons and $1$ for fermions. $m_i$ and $g_i$ are the mass and 
degeneracy factor of the $i$th hadron and $B_i$, $Q_i$ and $S_i$ are its conserved 
charges, namely baryon number, electric charge and strangeness respectively. Here 
$K_2$ is the Bessel function of the second kind. The total yield of the $i$th 
hadron $N^{\text{t}}_i$ comprise of the primordial yield as well as feeddown from 
heavier resonances that decay to it
\beq
 N_i^{\text{t}} = N_i^{\text{p}} + \sum_j N_j^{\text{p}}\times \text{B.R.}_{j\rightarrow i}
 \label{eq.totyield}
\eeq
where $\text{B.R.}_{j\rightarrow i}$ is the branching ratio of the channel in which the 
$j$th hadron decays to the $i$th hadron taken from P.D.G.~\cite{Beringer:1900zz}. In 1CFO, 
there is a single chemical freezeout surface and hence $T_i\l\sNN\r=T\l\sNN\r$ for all 
hadrons. Similarly, $V_i$, ${\mu_B}_i$, ${\mu_Q}_i$ and ${\mu_S}_i$ are same for all hadrons. 
In 2CFO, all strange hadrons and those with hidden strangeness freezeout at the same surface 
while the rest of the non strange hadrons freezeout at a separate surface. Thus 
$T_i=T_{\text{ns}}$ for all non strange hadrons while $T_i=T_{\text{s}}$ for all strange 
hadrons and those with hidden strangeness content. Volume and chemical potentials are also 
treated similarly. Within this framework, hadron yields were fitted and thermal paramters 
extracted for $\sNN=6.27-2700$ GeV in Ref.~\cite{Chatterjee:2013yga}. While the extracted 
fugacity factors for both the surfaces are similar with the exception at low beam energies 
$\sNN < 10$ GeV, the temperature and volume parameters clearly signal separation of CFO for 
non strange and strange hadrons at all energies~\cite{Chatterjee:2013yga}.
 
\subsection{COALESCENCE MODEL}
\label{subsec.2CFOCoalescence}

Within this picture, the nuclei are modelled to form by coalescence of hadrons near the KFO 
surface. This is usually expressed in terms of the invariant coalescence factor $B_A$
\beqa
E_A\frac{d^3N_A}{d^3P_A}&=&B_A\l E_p\frac{d^3N_p}{d^3P_p}\r^{Z}\l E_n\frac{d^3N_n}{d^3P_n}\r^{A-Z}
\label{eq.coalescence1}
\eeqa
where $A$ and $Z$ are the mass number and atomic number of the nuclei respectively. Depending 
on the choice of $B_A$, i.e. the mechanism in which one takes into account the hadronic 
correlations in the phase space at the final stages of the fireball evolution near KFO, 
there are various variants of the coalescence model~\cite{Butler:1963pp,Schwarzschild:1963zz,
Gutbrod:1988gt,Sato:1981ez,Scheibl:1998tk,Nagle:1996vp,Llope:1995zz,Danielewicz:1991dh,
Dover:1991zn,Csernai:1986qf,Gyulassy:1982pe,Remler:1981du,Zhang:2009ba,Steinheimer:2012tb}. 

A similar relation as Eq.~\ref{eq.coalescence1} at the level of yields can be written down
\beqa
N_A&=&C_A\l N_p\r^{Z}\l N_n\r^{A-Z}
\label{eq.coalescence2}
\eeqa
where the details of phase space correlations are encoded into $C_A$. Thus, ratio of anti 
nuclei to nuclei yields can be expressed in terms of ratios of their corresponding constituent 
hadrons
\beqa
  \frac{\overline{{N}_{A}}}{{N}_{A}}&=&
  C_{\overline{A}A}\l N_{\bar{p}}/N_p\r^Z\l N_{\bar{n}}/N_n\r^{\l A-Z\r}\label{eq.coalescenceratio1}\\
  &\sim&C_{\overline{A}A}\l N_{\bar{p}}/N_p\r^A\label{eq.coalescenceratio2}
\eeqa
where $C_{\overline{A}A}={C_{\overline{A}}}/{C_{A}}$. From Eq.~\ref{eq.coalescenceratio1} to 
Eq.~\ref{eq.coalescenceratio2} we assume $N_p\sim N_n$. Although Eqs.~\ref{eq.coalescence1}, 
\ref{eq.coalescence2}, \ref{eq.coalescenceratio1} and \ref{eq.coalescenceratio2} have been 
written for a nuclei containing only non strange hadrons like neutrons and protons as 
constituents, similar relation can be also written for hypernuclei containing strange hadrons 
like $\Lambda$. 
 
Thus nuclei production in the coalescence model is a combination of two distinct physics issues: 
(a) the physics of $C_{\overline{A}A}$ which is a subject of intense current research and an
agreement over its correct interpretation is yet to be reached \cite{Scheibl:1998tk}. This is 
related to the correlation effects in the phase space that exist between the constituent hadrons 
at the time of the KFO, and (b) the abundances of the constituent hadrons at the time of KFO which 
is already fixed at the CFO surface obtained from fits to the hadron yields. In this paper we are 
interested in the role played by the latter physics in determining the nuclei yields. Hence we will 
consider the simplest version of the coalescence model in which we will take 
$C_{\overline{A}A}=1$~\cite{Cleymans:2011pe}. This will be sufficient for our purpose to demonstrate 
the dependence of the nuclei production on the CFO scheme.
\section{Results}
\label{sec.results}

We will now present our results on the production of light nuclei and compare between 1CFO and
2CFO. Except pions, for all other hadrons and light nuclei, $m/T>>1$ and hence we can keep only 
the $l=1$ term in Eq.~\ref{eq.primyield} which is the Boltzmann approximation to get
\beqa
  N_i^{\text{p}}&=&\frac{V_iT_i}{\pi^2}g_im_i^2K_2\l m_i/T_i\r\times
  \exp\l \l B_i{\mu_B}_i+Q_i{\mu_Q}_i+S_i{\mu_S}_i\r/T_i\r\label{eq.Boltzprimyield}
\eeqa 
Now using the asymptotic expansion $K_{n}\l z\r\sim\sqrt{\frac{\pi}{2z}}\exp\l-z\r$ and 
neglecting the resonance feeddown, we have for ratio
\beqa
  N_i^{\text{t}}/N_j^{\text{t}}&=&
  \frac{g_iV_i}{g_jV_j}\l\frac{T_im_i}{T_jm_j}\r^{3/2}
  \exp\l m_j/T_j-m_i/T_i\r\exp\l B_i{\mu_B}_i/T_i-B_j{\mu_B}_j/T_j\r\times\nn\\
  &&\exp\l Q_i{\mu_Q}_i/T_i-Q_j{\mu_Q}_j/T_j\r\exp\l S_i{\mu_S}_i/T_i-S_j{\mu_S}_j/T_j\r\label{eq.ratio}
\eeqa
From Eq.~\ref{eq.ratio}, we can write for strange to non strange particle ratios in the 
thermal model
\beqa
  \l N_i^{\text{t}}/N_j^{\text{t}}\r^{\text{th}}&=&
  \exp\l S\mu_S/T_s \r\frac{g_iV_{s}}{g_jV_{ns}}\l\frac{T_sm_i}{T_{ns}m_{j}}\r^{3/2}
  \exp\l m_j/T_{ns}-m_i/T_s\r\exp\l {\mu_B}_s/T_s-{\mu_B}_{ns}/T_{ns}\r\label{eq.stons}
\eeqa
Now lets turn our attention to the coalescence model. For definiteness, we first look at 
${}^3_{\Lambda}\text{H}/{}^3\text{He}$. Modifying Eq.~\ref{eq.coalescenceratio2} suitably 
to take care of the $\Lambda$ in ${}^3_{\Lambda}\text{H}$, we get for the ratio 
${}^3_{\Lambda}\text{H}/{}^3\text{He}$ in the coalescence model
\beqa
\l {}^3_{\Lambda}\text{H}/{}^3\text{He}\r^{\text{c}}&=&\l\Lambda/\text{p}\r^{\text{th}}\label{eq.stonsc}
\eeqa

Similarly, for $\overline{{}^3_{\Lambda}\text{H}/{}^3\text{He}}$ we get
\beqa
  \l\overline{{}^3_{\Lambda}\text{H}/{}^3\text{He}}\r^{\text{c}}&=&
  \l\overline{\Lambda/\text{p}}\r^{\text{th}}\label{eq.antistonsc}
\eeqa

\begin{figure}[htb]
\begin{center}
  \includegraphics[scale=0.6]{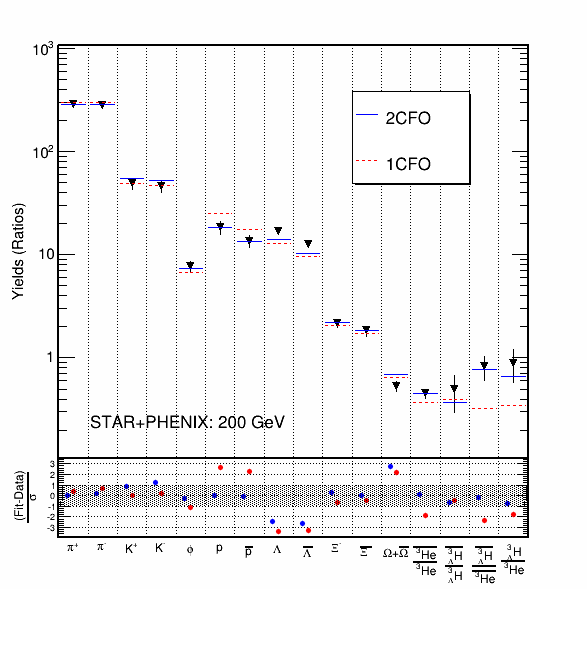}
\caption{(Color online) Thermal model production of hadrons and nuclei at $\sNN=200$ GeV. 
Here $\sigma$ is the error in the data. The solid inverted triangle 
represent the experimentally measured ratios~\cite{Abelev:2008aa,Abelev:2008ab,Adams:2004ux,
Adams:2006ke,Abelev:2010rv}. Only the data of hadrons were used to extract the thermal 
parameters.}
\label{fig.STARyield}
\end{center}
\end{figure}

Here $\l {}^3_{\Lambda}\text{H}/{}^3\text{He}\r^{\text{c}}$ and 
$\l\overline{{}^3_{\Lambda}\text{H}/{}^3\text{He}}\r^{\text{c}}$ refer to the nuclei ratios 
$\l {}^3_{\Lambda}\text{H}/{}^3\text{He}\r$ and $\l\overline{{}^3_{\Lambda}\text{H}/{}^3\text{He}}\r$ 
respectively in the coalescence model. Thus, from Eqs.~\ref{eq.stons}, \ref{eq.stonsc} and 
\ref{eq.antistonsc} we conclude that both in thermal as well as coalescence models when we consider 
strange to non strange light nuclei ratios, there is an additional prefactor in 2CFO depending on 
the different freezeout volumes and temperatures of the strange and non strange hadrons apart 
from the usual Boltzmann and fugacity factors that arise in 1CFO. This is a very general result in 
2CFO and is also true for strange to non strange hadron ratios like $\Lambda/$p. This makes such 
ratios quite sensitive probes to 2CFO. As we will see later in Fig.~\ref{fig.STARyield}, this leads 
to agreement between data and 2CFO prediction of the ratios $\l {}^3_{\Lambda}\text{H}/{}^3\text{He}\r$ 
and $\l\overline{{}^3_{\Lambda}\text{H}/{}^3\text{He}}\r$ which earlier 
studies~\cite{Andronic:2010qu,Cleymans:2011pe} based on 1CFO failed to explain. This is the main 
result of our paper.

\begin{figure}[htb]
\begin{center}
  \scalebox{0.35}{\includegraphics{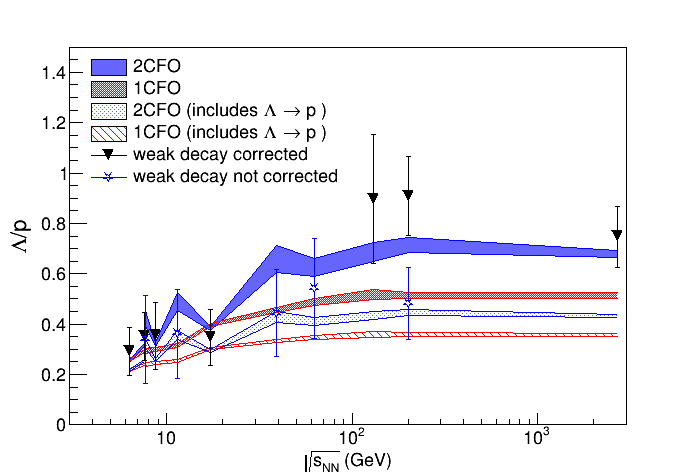}}
  \scalebox{0.35}{\includegraphics{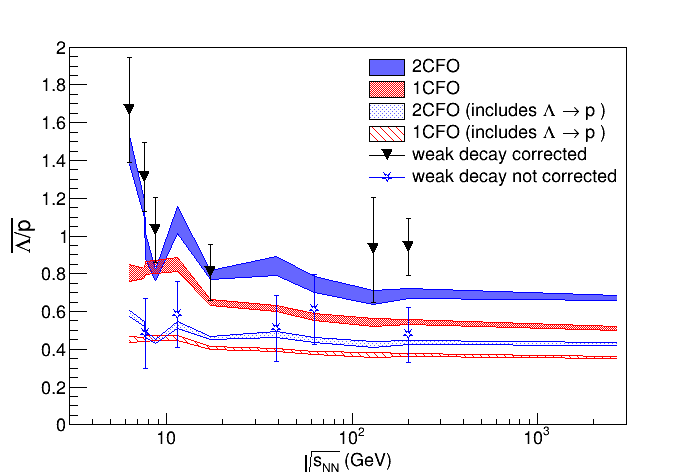}}\\
  \includegraphics[scale=0.35]{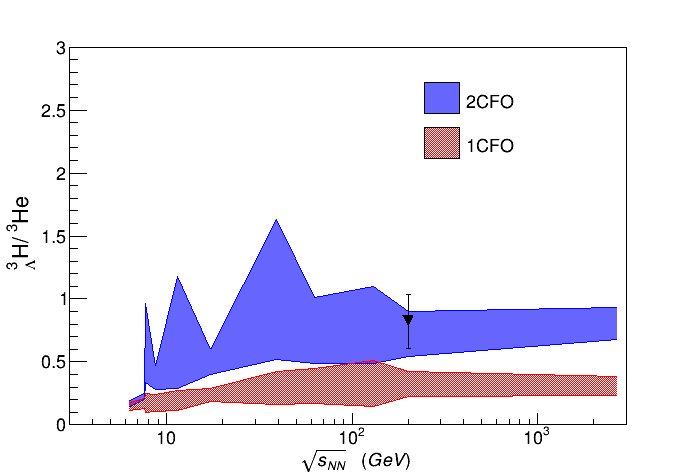}
  \includegraphics[scale=0.35]{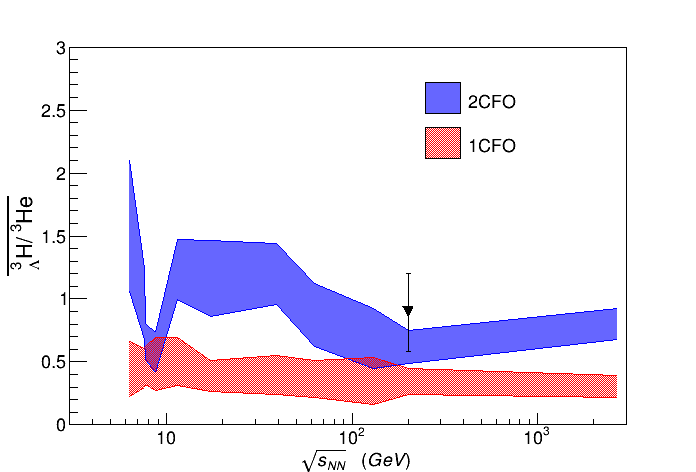}\\
\caption{(Color online) Plots of ratios of strange to non strange particles versus $\sNN$. 
The blue bands correspond to 2CFO and red bands to 1CFO. The width of the bands reflect 
the uncertainties associated with the ratios extracted in thermal model. The solid inverted 
triangle and open stars represent the experimentally measured ratios~\cite{Alt:2007aa,
Alt:2005gr,Alt:2008qm,Alt:2008iv,Afanasiev:2002mx,Alt:2006dk,Alt:2004kq,Abelev:2008aa,
Abelev:2008ab,Aggarwal:2010ig,Adler:2002xv,Adams:2003fy,Adams:2004ux,Adams:2006ke,
Adcox:2003nr,Adcox:2002au,Adler:2003cb,Milano:2013sza,Abelev:2010rv}.}
\label{fig.unlikeflavor}
\end{center}
\end{figure}

From Eqs.~\ref{eq.stons}, \ref{eq.stonsc} and \ref{eq.antistonsc} we may write
\beqa
  \l {}^3_{\Lambda}\text{H}/{}^3\text{He}\r^{\text{th}}/
  \l {}^3_{\Lambda}\text{H}/{}^3\text{He}\r^{\text{c}}&=&
  \l\frac{ m_{{}^3_{\Lambda}\text{H}} \ m_{\text{p}}}
  {m_{\Lambda}\ m_{{}^3\text{He}}}\r^{3/2}
  \exp\l \l m_{{}^3\text{He}}-m_{\text{p}}\r/T_{ns}-
  \l m_{{}^3_{\Lambda}\text{H}}-m_{\Lambda}\r/T_{s} \r\label{eq.thvsc1}\\
  &\simeq&\l 1/3+2/3\l m_{\text{p}}/m_{\Lambda}\r\r^{3/2}\exp\l \frac{2m_p
  \l 1-\frac{T_{ns}}{T_s}\r}{T_{ns}}\r\label{eq.thvsc2}
\eeqa
From Eq.~\ref{eq.thvsc1} to~\ref{eq.thvsc2} we have used the fact that binding 
energy of ${}^3_{\Lambda}\text{H}$ and ${}^3\text{He}$ can be neglected compared 
to the nuclei masses as well as the fireball temperatures at the two freezeout 
surfaces. Here we would like to make few observations. If we put $T_{ns}=T_s=T$, 
we recover the result as expected in 1CFO~\cite{Cleymans:2011pe}. Secondly, since 
the factor $\l 1/3+2/3\l m_{\text{p}}/m_{\Lambda}\r\r^{3/2}\sim0.85$, in case of 
1CFO the thermal value would be always smaller than that in the coalescence 
case~\cite{Cleymans:2011pe}. However in 2CFO because of the additional 
exponential factor, depending on the value of $T_{ns}/T_s$ as extracted from fits 
to hadron yields, the thermal value could be lesser or more than the coalescence 
value. In Fig.~\ref{fig.unlikeflavor}, we have  plotted the particle ratios 
$\l{}^3_{\Lambda}\text{H}/{}^3\text{He}\r^{\text{th}}$ and 
$\l\overline{{}^3_{\Lambda}\text{H}/{}^3\text{He}}\r^{\text{th}}$ as well as 
$\l\Lambda/\text{p}\r^{\text{th}}$ $\l=
\l{}^3_{\Lambda}\text{H}/{}^3\text{He}\r^{\text{c}}\r$ and 
$\l\overline{\Lambda/\text{p}}\r^{\text{th}}$ $\l=
\l\overline{{}^3_{\Lambda}\text{H}/{}^3\text{He}}\r^{\text{c}}\r$
versus $\sNN$. We find that within the uncertainties due to
the errors in the fit parameters, thermal and coalescence values for the ratios 
agree. Thus as in 1CFO, it is not possible to discriminate between thermal and 
coalescence mechanism of nuclei production in 2CFO. However, the bands for 1CFO 
and 2CFO are quite distinct and hence they are good candidates to distinguish 
between the different freezeout scenarios. Currently the bands are based on the 
errors in the extraction of the thermal parameters from hadron 
yields~\cite{Chatterjee:2013yga}, many of which are based on preliminary data 
from Beam Energy Scan (BES) programme of STAR~\cite{Das:2012yq}. We expect the 
bands to get narrower in the future with the availability of final data from BES. 
It is to be noted that the weak decay contribution from $\Lambda$ to proton is 
treated differently for different experiments. While the STAR data for proton 
includes the weak decay contribution from $\Lambda$~\cite{Abelev:2008aa,
Aggarwal:2010ig,Adler:2002xv,Adams:2003fy,Adams:2004ux,Adams:2006ke,Abelev:2008ab}, 
data from NA49~\cite{Alt:2007aa,Alt:2005gr,Alt:2008qm,Alt:2008iv,Afanasiev:2002mx,
Alt:2006dk,Alt:2004kq}, PHENIX~\cite{Adcox:2003nr,Adcox:2002au,Adler:2003cb} and 
ALICE~\cite{Milano:2013sza} are corrected from such decays. We note from 
Fig.~\ref{fig.unlikeflavor} that broadly speaking, while  $\Lambda/$p and 
${}^3_{\Lambda}\text{H}/{}^3\text{He}$ 
rise and then saturate with increasing $\sNN$, $\overline{\Lambda/\text{p}}$ and 
$\overline{{}^3_{\Lambda}\text{H}/{}^3\text{He}}$ first fall and then stay unchanged 
with increasing $\sNN$ (at lower energies they are expected to go to zero as $\Lambda$
($\overline{\Lambda}$) is heavier than $p$ ($\overline{p}$)). Thus the different shapes of 
$\Lambda/$p ( ${}^3_{\Lambda}\text{H}/{}^3\text{He}$) and $\overline{\Lambda/\text{p}}$ 
($\overline{{}^3_{\Lambda}\text{H}/{}^3\text{He}}$) can be attributed to
the strange fugacity factor $\exp\l S\mu_S/T\r$. With decreasing $\sNN$ as $\mu_S$ 
increases, the strange fugacity factor enhances the difference between 1CFO and 
2CFO for $\overline{\Lambda/\text{p}}$ since $\overline{\Lambda}$ has $S=1$. 
The effect of the strange fugacity factor is just the opposite on $\Lambda/$p as 
$\Lambda$ has $S=-1$. As seen from Fig.~\ref{fig.unlikeflavor}, at the FAIR energy 
range this makes $\overline{\Lambda/\text{p}}$ and 
$\overline{{}^3_{\Lambda}\text{H}/{}^3\text{He}}$ excellent candidates to distinguish 
between 1CFO and 2CFO freezeout scenarios. However, it has to be kept in mind that 
at the FAIR the production of antibaryons will be highly suppressed due to large 
baryon chemical potential.

\begin{figure}[htb]
\begin{center}
  \includegraphics[scale=0.35]{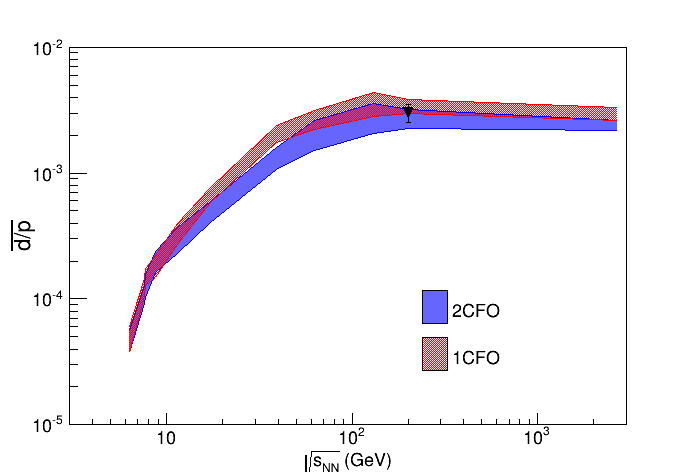}
  \includegraphics[scale=0.35]{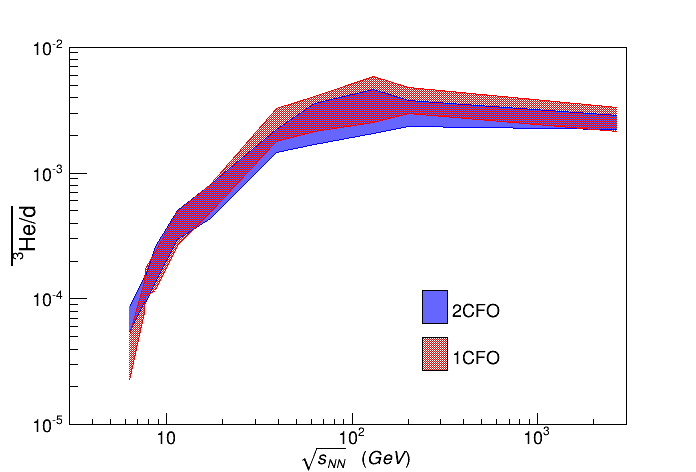}\\
  \scalebox{0.35}{\includegraphics{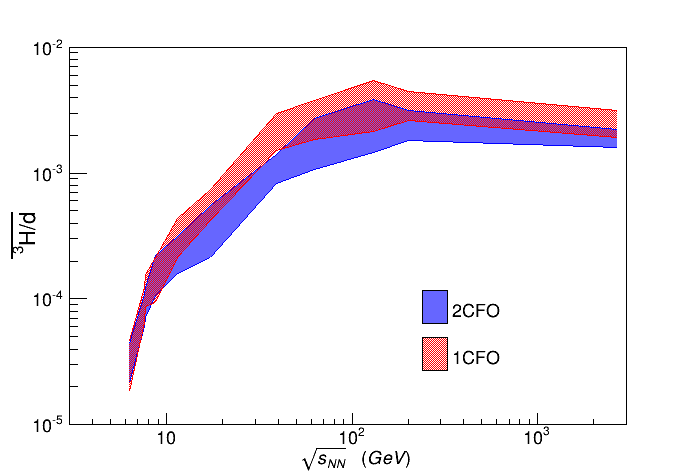}}
  \scalebox{0.35}{\includegraphics{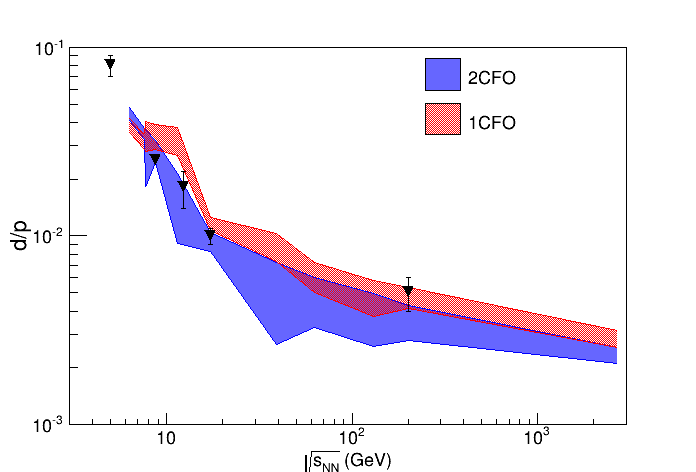}}\\
  \scalebox{0.35}{\includegraphics{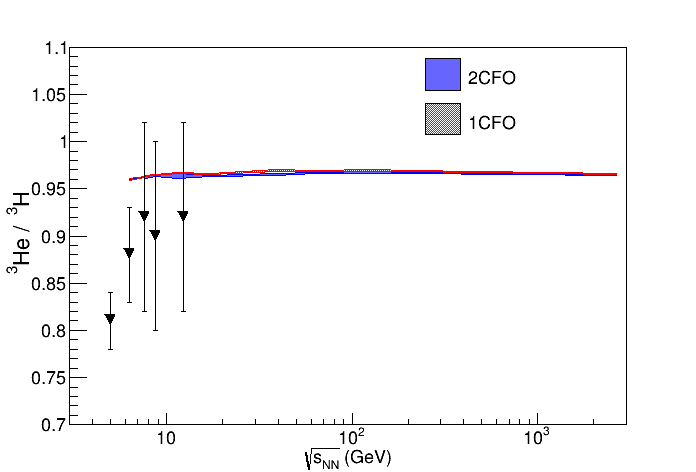}}
  \includegraphics[scale=0.35]{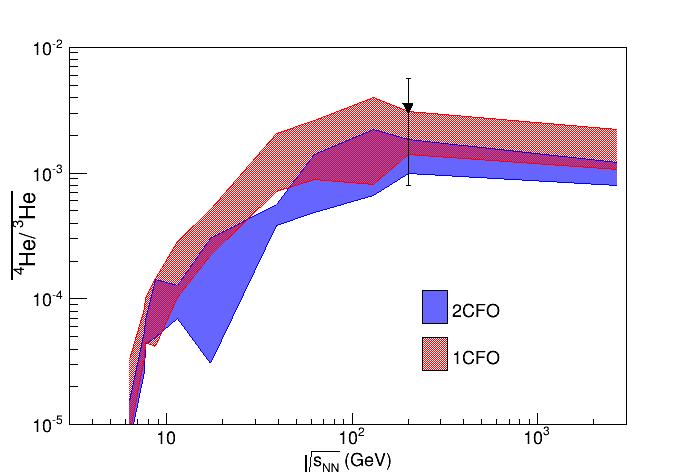}\\
\caption{(Color online) Plots of ratios of non strange particles versus $\sNN$. The blue 
bands correspond to 2CFO and red bands to 1CFO. The width of the bands reflect the 
uncertainties associated with the ratios extracted in thermal model. The solid inverted 
triangles represent the experimentally measured ratios~\cite{Alt:2007aa,
Alt:2005gr,Alt:2008qm,Alt:2008iv,Afanasiev:2002mx,Alt:2006dk,Alt:2004kq,Abelev:2008aa,
Abelev:2008ab,Aggarwal:2010ig,Adler:2002xv,Adams:2003fy,Adams:2004ux,Adams:2006ke,
Adcox:2003nr,Adcox:2002au,Adler:2003cb,Milano:2013sza,Abelev:2010rv,Agakishiev:2011ib}.}
\label{fig.sameflavor}
\end{center}
\end{figure}

Eq.~\ref{eq.ratio} gets further simplified if we consider ratios of particles with 
same flavor like $d/p$, $^3_\Lambda\text{H/}\Lambda$ or those involving antiparticles 
like $\overline{\text{d/p}}$, $\overline{{}^4\text{He}/{}^3\text{He}}$ etc.
\beqa
  N^t_i/N^t_j&=&
  \l\frac{g_i}{g_j}\r\l\frac{m_i}{m_j}\r^{3/2}
  \exp\l\l\l m_j-m_i\r+\l B_i-B_j\r\mu_B+\l Q_i-Q_j\r\mu_Q+
  \l S_i-S_j\r\mu_S\r/T\r\label{eq.particleratiosameflavor}
\eeqa
In this case the prefactor that arose in the earlier case drops out. Hence we expect 
similar predictions for 1CFO and 2CFO. We have plotted some of these ratios in 
Fig.~\ref{fig.sameflavor}. As expected the bands for 1CFO and 2CFO almost overlap across 
the entire beam energies.

\begin{figure}[htb]
\begin{center}
  \includegraphics[scale=0.5]{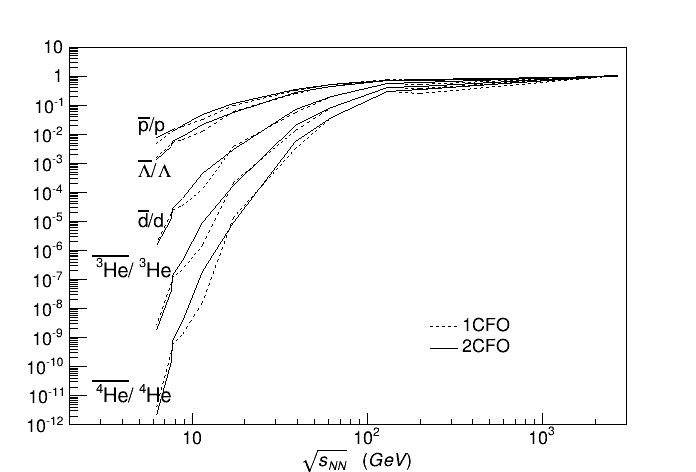}
\caption{Few antiparticle to particle ratios of light nuclei. As argued in the text, 
these ratios are not sensitive to the different freezeout mechanisms.}
\label{fig.antiparticletoparticle}
\end{center}
\end{figure}

For anti-particle to particle ratios, Eq.~\ref{eq.particleratiosameflavor} simplifies 
even further. Except the fugacities, all other factors drop out. Thus from 
Eqs.~\ref{eq.coalescenceratio2} and \ref{eq.particleratiosameflavor} we may write
\beqa
  \l \overline{N^t_i}/N^t_i\r^{\text{c}}=\l \overline{N^t_i}/N^t_i\r^{\text{th}}
  &=&\exp\l-2\l B_i{\mu_B}_i+Q_i{\mu_Q}_i+S_i{\mu_S}_i\r/T\r\label{eq.antiparticlebyparticle}
\eeqa
Thus one can directly extract the fugacity factors using such ratios. Since nuclei 
have $B>1$, anti nuclei to nuclei ratios are even more sensitive to the baryon fugacity 
factor compared to that of hadrons~\cite{Andronic:2010qu,Cleymans:2011pe}. However as 
expected from Eq.~\ref{eq.antiparticlebyparticle},
these ratios are not sensitive to the different freezeout schemes. We have plotted in 
Fig.~\ref{fig.antiparticletoparticle} a few anti-particle to particle ratios, namely 
$\overline{\text{p}}/\text{p}$, $\overline{\Lambda}/\Lambda$, $\overline{\text{d}}/\text{d}$, 
$\overline{{}^3\text{He}}/{}^3\text{He}$ and $\overline{{}^4\text{He}}/{}^4\text{He}$ versus 
$\sNN$. The error bands have not been shown due to clarity.

So far we have analysed several light nuclei ratios as function of $\sNN$. We have plotted
these ratios for 1CFO and 2CFO freezeout schemes. We argued as well as demonstrated that 
ratios of hadrons and nuclei with same flavor are insensitive to these different freezeout 
schemes. Similarly, while ratios of anti nuclei to nuclei are known to be very sensitive to 
the baryon fugacity factor, they are not suitable to distinguish between the different 
freezeout schemes, 1CFO and 2CFO. However, we showed that ratios of strange to non strange 
hadrons and nuclei can discriminate between the different freezeout schemes. Moreover, we 
argued that even if we change the production mechanism from thermal to a simple coalescence, 
the above statement on the sensitivity of the various types of nuclei ratios on the CFO scheme 
remain true. This completes our study of the light nuclei production for different freezeout 
schemes over a broad range of $\sNN$. Now we will focus on the following beam energies: 
$\sNN=200$ and $2700$ GeV. At these energies, data for a large number of hadrons as well as 
light nuclei including strange nuclei like ${}^3_{\Lambda}\text{H}$ are available inviting 
for a more detailed inspection.

\begin{figure}[htb]
\begin{center}
  \includegraphics[scale=0.6]{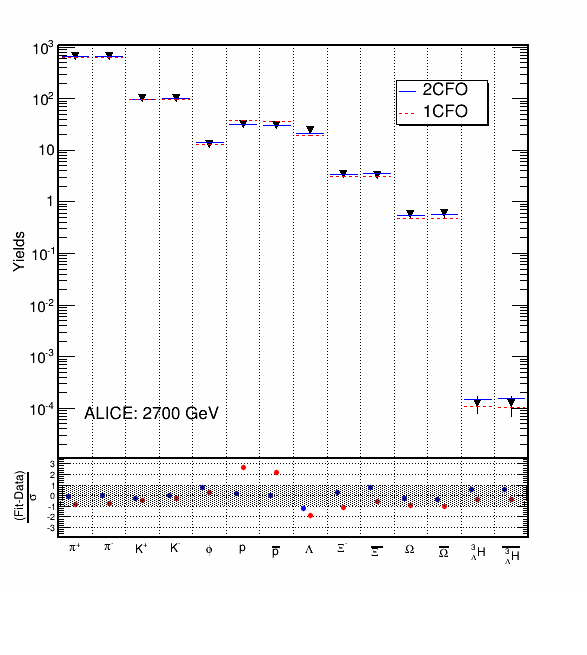}
\caption{(Color online) Thermal model production of hadrons and nuclei at $\sNN=2700$ GeV. 
Here $\sigma$ is the error in the data. The solid inverted triangle 
represent the experimentally measured ratios~\cite{Abelev:2013vea,Abelev:2013xaa,
ABELEV:2013zaa,alice1}. Only the data of hadrons were used to extract the thermal parameters.}
\label{fig.ALICEyield}
\end{center}
\end{figure}

In Fig.~\ref{fig.STARyield}, we have shown the results obtained at $\sNN=200$ GeV. The 
fits of the results as shown in Table~\ref{tab.fit} are found to change little when we 
include even the light nuclei in the fits. It is found that 2CFO is able to correctly 
predict ${}^3_{\Lambda}\text{H}/{}^3\text{He}$ ratio while 1CFO doesn't.  There is 
mismatch with data in the strange baryon sector. The cause for this could be twofold: 
firstly a proper treatment of the weak decays in the STAR data set is necessary before 
anything conclusive can be said. Secondly, this could be a hint for further possible 
structures in the freezeout mechanism, for example, an early freezeout of the strange 
baryons. It is to be noted that including the strangeness undersaturation factor 
$\gamma_S$ improves the agreement with data considerably in this case. The issue of the 
weak decays is much better addressed in ALICE. We have shown the fits to the latest 
$\l0-10\r$ $\%$ ALICE data~\cite{Abelev:2013vea,Abelev:2013xaa,ABELEV:2013zaa,alice1} in 
Fig.~\ref{fig.ALICEyield}. The 2CFO mechanism seems to describe the data for all the 
particles including light nuclei unlike the 1CFO where the protons are not described well. 
In the future, the LHC experiment is also expected to produce more high precision data on 
nuclei which could provide more insight into the freezeout mechanism.

\begin{table*}[bt]
\begin{center}
\begin{tabular}{|r@{.}l|c|c|c|c|c|c|c|c|c|}
\hline
 \multicolumn{2}{|c|}{$\sqrt{s_{NN}}$} & nuclei &
 $10^4V_S$ & $10^4V_{NS}$ &
 $T_S$ & $T_{NS}$ &
 $\mu_S$ & $\mu_{NS}$ & $\chi^2/N_{df}$ \\
 \multicolumn{2}{|c|}{(GeV)} & fitted &
 (MeV$^{-3}$) & (MeV$^{-3}$) & (MeV) & (MeV) & (MeV) & (MeV) & \\
\hline
 200& &No   
         &2.2 (0.4)&2.8 (0.8)&164 (3)&155 (6)& 31 (11)& 22 (16)&23/6\\
 200& &Yes   
         &2.3 (0.4)&2.6 (0.8)&163 (3)&155 (6)& 27 (11)& 23 (16)&24/10\\        
         \hline\hline
2700& &No
         &5.5 (0.6)&9.7 (0.8)&158 (3)&145 (3)& 0 (12)& 0  (7)&3.1/6\\
\hline
\end{tabular}
\end{center}
\caption{The freezeout parameters in 2CFO at top RHIC energy of $\sNN=200$ GeV and LHC 
energy of $\sNN=2700$ GeV.}
\label{tab.fit}\end{table*}

\section{Summary and Conclusion}
\label{sec.conc}
We have studied the production of light nuclei in heavy ion collisions within thermal and 
simple coalescence models in the light of multiple freezeout scenarios. We argued and 
showed that irrespective of the production mechanism, while ratios of same flavor nuclei 
(non strange or strange) are insensitive to the chemical freezeout scheme chosen, mixed 
ratios i.e. ratios of unlike flavor nuclei can probe the details of the chemical freezeout. 
This is in general true for hadrons also. Particularly at $\sNN=200$ GeV, hitherto 
unexplained nuclei ratios  ${}^3_{\Lambda}\text{H}/{}^3\text{He}$ and 
$\overline{{}^3_{\Lambda}\text{H}/{}^3\text{He}}$ within the assumption of a single chemical 
freezeout surface, is found to agree with model predictions when we modify the chemical 
freezeout scheme such that strange and non strange hadrons freezeout separately. In the future, 
it would be interesting to investigate the effect of including in HRG yet undiscovered resonances 
that are predicted by lattice QCD as well as quark models on the above ratios~\cite{Bazavov:2014xya}.

\section{Acknowledgement}
We acknowledge helpful discussions on nuclei production with Sourendu Gupta. 
SC acknowledges financial support from DST SwarnaJayanti project of BM. This work is 
also supported by DAE-SRC project.

\bibliographystyle{apsrev4-1}
\bibliography{Lightnuclei}

\end{document}